\documentclass[%
 reprint,
superscriptaddress,
showpacs,preprintnumbers,
 amsmath,amssymb,
 aps,
prb,
]{revtex4-1}
\usepackage{graphicx}
\usepackage{dcolumn}
\usepackage{bm}

\begin{document}

\title{Gate-control of spin-motive force and spin-torque in Rashba SOC systems}

\author{Cong Son Ho}
 \affiliation{
Department of Electrical and Computer Engineering, National University of Singapore,4 Engineering Drive 3, Singapore 117576, Singapore.
}
\author{Mansoor B. A. Jalil}%
 \email{elembaj@nus.edu.sg}
\affiliation{
Department of Electrical and Computer Engineering, National University of Singapore,4 Engineering Drive 3, Singapore 117576, Singapore.
}
\author{Seng Ghee Tan}
\affiliation{
Department of Electrical and Computer Engineering, National University of Singapore,4 Engineering Drive 3, Singapore 117576, Singapore.
}
\affiliation{
Data Storage Institute, Agency for Science, Technology and Research (A*STAR),
DSI Building, 5 Engineering Drive 1,Singapore 117608, Singapore.
}

\date{\today}

\begin{abstract}
The introduction of a strong Rashba spin orbit coupling (SOC) has been predicted to enhance the spin motive force (SMF) [see Phys. Rev. Lett. {\bf 108}, 217202 (2012)]. In this work, we predict further enhancement of the SMF by time modulation of the Rashba coupling $\alpha_R$, which induces an additional electric field $E^R_d={\dot \alpha_R} m_e/e\hbar({\hat z}\times {\mathbf m})$. When the modulation frequency is higher than the magnetization precessing frequency, the amplitude of this field is significantly larger than previously predicted results. Correspondingly, the spin torque on the magnetization is also effectively enhanced. We also suggest a biasing scheme to achieve rectification of SMF, {\it i.e.}, by application of a square wave voltage at the resonant frequency. Finally, we numerically estimate the resulting spin torque field arising from a Gaussian pulse time modulation of $\alpha_R$.
\end{abstract}

\maketitle

\section{Introduction}

One of the most important subjects of current study in spintronics is the manipulation of spin and magnetization. A spin current can modify the dynamics of magnetization through spin-transfer torque (STT); \cite{PhysRevB.54.9353,Kiselev03,Lee04} conversely the dynamics of magnetization can in turn modify the spin dynamics and may generate a spin current through spin motive force (SMF)\cite{PhysRevB.33.1572,0022-3719-20-7-003,PhysRevLett.98.246601,PhysRevLett.102.086601} and spin pumping. \cite{PhysRevLett.88.117601,PhysRevB.76.184434,PhysRevB.77.014409} Despite the mutual connection between STT and SMF, the latter has been shown to be inefficient in generating the spin current due to its weak magnitude,\cite{PhysRevLett.102.067201,Ohe09,PhysRevLett.107.236602} hence the on-going research on the means of obtaining a large SMF.

Recently, Kim {\it et al.}\cite{Kim12} predicted that in systems with large Rashba spin-orbit coupling,\cite{Mihai10,PhysRevB.77.214429,Pi10,Ishizaka11,Bahramy12} the induced SMF is more than an order of magnitude larger than the conventional SMF. In such systems, the Rashba-induced SMF (RSMF) can generate large spin current, which in turn can  significantly enhance the STT. However, the requirement of the strong Rashba coupling appears to be a limitation of the prediction in the sense that it is hard to increase the RSOC strength in order to enhance the RSMF. In systems with typical values of the Rashba coupling \cite{PhysRevLett.91.056602}, the conventional SMF and Rashba-induced SMF are almost comparable.  Moreover, in contrast to the conventional SMF which can be made time-independent (dc), the Rashba-induced SMF is not rectified, {\it i.e.}, it oscillates with time (ac). Therefore, enhancement and modulation of the SMF over a broader range of Rashba SOC strength is still essential.

In this study, we propose a method that can significantly enhance the Rashba SOC spin-motive force, in which a strong Rashba coupling is not prerequisite. The idea is based on the fact that the Rashba coupling can be modulated in time\cite{Datta90,Nit97,Grundler00,Caviglia10,PhysRevB.89.220409} by applying some AC gate voltage ($\omega_R$). In such systems, the dynamics of both spin and magnetization are controlled by either the Rashba amplitude and/or Rashba modulation frequency. While the former is a material-dependent parameter limited to its reported values, the latter can be easily modified by external means. We showed that in the presence Rashba coupling that is sinusoidally modulated, the resulting RSMF is enhanced with increasing modulation frequency. If $\omega_R>\omega_0$, with $\omega_0$ being the angular frequency of the magnetization dynamics, the RSMF can exceed that induced by a constant Rashba SOC. \cite{Kim12} Therefore, the generated spin current and the corresponding STT can be enhanced by the time modulation of the Rashba coupling. Moreover, the RSMF can also be rectified by applying a square-wave gate voltage, {\it i.e.}, the SMF is unidirectional, rendering it more useful for practical spintronics.

\section{Spin motive force}
The conduction electron in a magnetization texture with Rashba SOC can be described by the following Hamiltonian:
\begin{equation}\label{eq1}
H=\frac{{\mathbf p}^2}{2m_e}-J_{\textrm ex}\left({\hat \sigma}\cdot{\mathbf m}\right)+H_R
\end{equation}
where $\hat\sigma$ is the vector of Pauli matrix, $m_e$ is the effective mass of electron, $J_{\textrm ex}$ is the exchange coupling, $\mathbf m=(\sin\theta \cos\phi,\sin\theta \sin\phi,\cos\theta)$ is the unit vector of the local magnetization, and $H_R=(\alpha_R/\hbar) {\hat \sigma}\cdot({\mathbf p}\times{\hat z})$ is the Rashba Hamiltonian, with $\alpha_R$ being the Rashba coupling.

To model the effect of the precessing magnetization and the Rashba SOC on the electron dynamics, we adopt the conventional method,\cite{0022-3719-20-7-003,Kim12} {\it i.e.}, by introducing an unitary transformation $U^\dagger=e^{i \theta \sigma_y/2} e^{i \phi \sigma_z/2}$. With this transformation, the above Hamiltonian becomes
\begin{equation}\label{eq2}
H'=\frac{({\mathbf p}+e{\mathbf A}' )^2}{2m_e }-J_{\textrm ex} \sigma_z-eA_0',
\end{equation}
in which ${\mathbf A'}$ is the vector potential and it is given by:
\begin{equation}\label{eq3}
{\mathbf A'}=-\frac{i\hbar}{e} U^\dagger\nabla U+U^\dagger {\mathbf A}^R U,
\end{equation}
with ${\mathbf A^R}=-(\alpha_R m_e)/e\hbar ({\hat \sigma}\times\hat z )$ is the non-Abelian gauge field associated with Rashba SOC, which is explicitly time-dependent, and
\begin{equation}\label{eq4}
A_0'=i\hbar/e U^\dagger \partial_t U.
\end{equation}
These gauge fields will induce an effective electric field as $\mathbf E=-\partial_t{\mathbf A'}-\nabla A_0'$. We assume that the exchange coupling is very strong so that the spin tends to align along $\mathbf m$ in the lab frame (along ${\mathbf m}'=(0,0,1)$ in the rotated frame). With this, the effective electric field reads as:
\begin{equation}\label{eq5}
{\mathbf E}^{\uparrow(\downarrow)}=\pm\left({{\mathbf E}^m+\mathbf E}^{R,0}+{\mathbf E}^{R,1}\right),
\end{equation}
where the $\pm$ is corresponding to the majority ($\uparrow$) and minority ($\downarrow$) electrons, respectively. In the above, the first term
\begin{equation}\label{eq6}
{E}_i^{m}={\frac{\hbar}{2e}} (\partial_t{\mathbf m}\times\partial_i{\mathbf m})\cdot{\mathbf m},
\end{equation}
is the conventional electric field induced by the variation of the magnetization pattern in space and time \cite{PhysRevLett.102.086601,PhysRevLett.98.246601} in the absence of SOC, and
\begin{equation}\label{eq7}
{E}_i^{R,0}= \alpha_R  \frac{m_e}{e\hbar} (\partial_t {\mathbf m}\times{\hat z})_i,
\end{equation}
which is scaled with the value of the Rashba coupling,\cite{Kim12} and the last term
\begin{equation}\label{eq8}
{E}_i^{R,1}={\dot\alpha_R}  \frac{m_e}{e\hbar} ({\mathbf m}\times{\hat z})_i,
\end{equation}
is an extra term being dependent on the time variation of the Rashba coupling. In previous works,\cite{Ho12, Ho13} we showed that in a time-dependent Rashba system, there is an effective field induced by the time-dependent gauge field, i.e, ${\mathcal E}={\dot\alpha_R}  \frac{m_e}{e\hbar} ({\hat\sigma}\times{\hat z})$. For strong exchange coupling, we have $\hat\sigma=\pm {\mathbf m}$, which recovers Eq.~\eqref{eq8}. Remarkably, we can obtain Eq.~\eqref{eq8} without the need of the unitary transformation, implying that this electric field can be generated in any magnetization pattern, which can even be static or uniform,\cite{PhysRevB.88.014430} and the driving force in this case is the gate-modulated Rashba SOC. Therefore, we can electrically control the generation of the total SMF via an applied gate voltage.

\section{Gate-controlled spin-motive force}
\begin{figure}
\centering
 \includegraphics[width=0.4\textwidth]{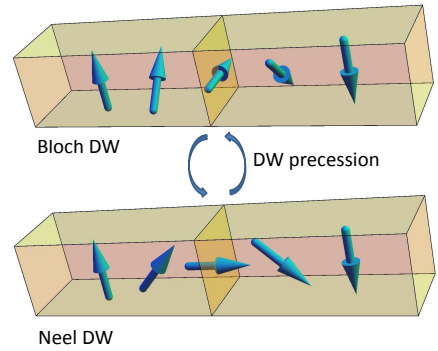}
\caption{The oscillation of the domain wall (DW): the DW is assumed to precess periodically between Block-like DW and Neel-like DW at frequency $\omega_0$.}
\label{Fig4_2}
\end{figure}

Consider a magnetization profile ${\mathbf m}[\theta({\mathbf{r}}),\phi(t)]$, where $\phi(t)=\omega_0 t$, and the spatial-dependence is implied by $\theta$ so that ${\vec{\nabla}}\theta\propto 1/L$, with $L$ being the characteristic length of the magnetic structure,\cite{Kim12} e.g., the domain wall width. This magnetization configuration can be made by assuming that the domain wall (DW) precesses periodically between the Bloch-like DW and Neel-like DW (see Fig.\ref{Fig4_2}). From now on, we only use this magnetization profile for simplicity. In this case, the above effective electric fields are explicitly given as
\begin{eqnarray}
{\mathbf E}^{m}&=&-{\frac{\hbar}{2e}} \omega_0 \sin\theta {\vec{\nabla}}\theta,\label{eq8a}\\
{\mathbf E}^{R,0}&=&- \frac{m_e}{e\hbar}\alpha_R \omega_0 \sin{\theta} {\mathbf n},\label{eq8b}\\
{\mathbf E}^{R,1}&=&- \frac{m_e}{e\hbar}{\dot \alpha_R} \sin{\theta} ({\mathbf n}\times{\hat{z}}),\label{eq8c}
\end{eqnarray}

\noindent where ${\mathbf {n}}=(\cos \omega_0 t,\sin\omega_0 t,0)$. It is obvious that the effective in Eq. \eqref{eq8a} is independent of time. Meanwhile, for a constant Rashba coupling, the field in Eq. \eqref{eq8b} oscillates with time. However, since these electric field components depend on both magnetization profile and the Rashba dynamics, we may rectify and enhance the spin motive force  by introducing a suitable AC gate modulation of the Rashba coupling. In following, we will show that by appropriately modulating the Rashba coupling, the electric field induced by RSOC can be significantly enhanced and rectified.

\subsection{Enhancement of SMF}

Consider a sinusoidal modulated Rashba coupling $\alpha_R(t)=\alpha_0 \cos(\omega_R t)$. In this case, the order of magnitude of the above electric field components are evaluated to be $E^m=\frac{\hbar\omega_0}{eL}\sin\theta, E^{R,0}= \alpha_0 \frac{m_e}{e\hbar} \omega_0 \sin\theta, E^{R,1}= \alpha_0 \frac{m_e}{e\hbar} \omega_R \sin\theta$. The ratio $E^{R,0}/E^m=\alpha_0 m_eL/\hbar^2$ shows that one can enhance the SMF in the presence of strong Rashba coupling of $\alpha>\hbar^2/m_eL$.\cite{Kim12} On the other hand, the ratio $E^{R,1}/E^{R,0}=\omega_R/\omega_0$ suggests that the additional SMF component due to the time-modulation of the Rashba coupling can have a comparable or larger magnitude if the Rashba modulation $\omega_R\ge\omega_0$. Thus, with time-modulation of the Rashba coupling, i) one obtains an additional component to enhance the SMF; ii) this enhancement is no longer just restricted by the requirement of strong Rashba coupling; and iii) the frequency of modulation $\omega_R$ provides another external parameter to control the size of the SMF. For example, in systems with Rashba coupling such as GaMnAs having $\alpha_R=10^{-11}~{\mathrm{eV m}}$,\cite{PhysRevLett.91.056602,PhysRevB.78.212405} and assuming $L=20$ nm, $\omega_0=2\pi\times 100~\mathrm{MHz}$, the conventional SMF signal comes up to $V(E^m)\approx 0.4~\mu\mathrm V$, meanwhile the RSOC SMF signal is $V(E^{R,0})\approx 1~\mu\mathrm V$, which just has the same order of magnitude as the former. However, upon modulating the Rashba coupling at a frequency of $\omega_R=2\pi\times $ GHz,\cite{Mal03,PhysRevB.78.245312} the new SMF signal can be further increased up to $V(E^{R,1})\approx 10~\mu\mathrm V$.  We shall now discuss the rectification of the RSMF signal which would render it more useful for practical spintronic applications.

\subsection{Rectification of SMF}
	
	\begin{figure}
	\includegraphics[width=0.4\textwidth]{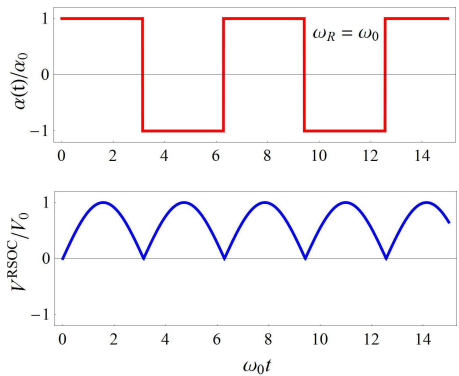}





 \caption{The rectification of the Rashba spin-orbit SMF by applying a square wave gate voltage at resonant frequency $\omega_R=\omega_0$. Here, $V_0=\frac{m_eL}{e\hbar}\alpha_0\omega_0$ is the amplitude of the SMF, where $L$ is the domain wall width. \label{Fig4_1}}
\end{figure}

In the previous section, we showed that the introduction of Rashba SOC can induce a large SMF. However, the Rashba-induced SMF generally oscillates with time, and so averages to zero over time. Here, we will show that, by applying a square-wave gate voltage, one can rectify one of the Rashba-induced SMF components.

Let us suppose the Rashba coupling is modulated as $\alpha_R(t)=\alpha_0 f(t)$, with $f(t)=\mathrm{sgn}(\sin\omega_R t)$ representing a square wave function [see Fig.\ref{Fig4_1}]. In this case, $E^{R,1}=0$ since $\dot\alpha_R=0$. On the other hand, by choosing $\omega_R=\omega_0$, the sign of $\alpha$ vary in step with the magnetization orientation, yielding $E^{R,0}$ with a constant sign. In this case, $E^{R,0}$ in Eq. \eqref{eq8b} becomes
\begin{equation}\label{eq9a}
{E}_{y}^{R,0}=\mp \frac{m_e}{e\hbar}\alpha_0 |\sin\omega_R t|.
\end{equation}
The above electric field and the corresponding SMF is unidirectional, although its value changes with time (pulsating) [see Fig. \ref{Fig4_1}]. Similarly, we can rectify the SMF component in the $x$-direction by applying a $\pi/2$ phase shift to the gate voltage, {\it i.e.}, $\alpha_R(t)\rightarrow\alpha_R(t+\pi/2\omega_R)$. This rectification effect is similar to that in spin-torque diode effect,\cite{ST.Diode05,ST.Diode13} where a dc voltage is generated when the frequency of the applied alternating current is resonant with the spin oscillations.

\section{Spin current-induced spin torque}

We have derived the effective electric fields generated by the dynamics of either the magnetization or the electron spin. These electric fields can drive a charge current and a spin current. In turn, the spin current induces a torque on the magnetization through spin-transfer mechanism \cite{PhysRevLett.102.086601}, while the charge current contributes a field-like torque.\cite{Tan07,Tan11,PhysRevB.78.212405,PhysRevB.79.094422, PhysRevB.77.214429,PhysRevB.80.094424}  Generally, the spin current $J_s$ and  charge current $J_e$ induced by the SMF is given by
\begin{eqnarray}
J_{s,i}=\frac{g\mu_B}{2e}\left(G^\uparrow E^\uparrow_i-G^\downarrow E^\downarrow_i\right)=\frac{g\mu_B G_0}{2e} E^\uparrow_i,\label{eq10}\\
J_{e,i}=\left(G^\uparrow E^\uparrow_i+G^\downarrow E^\downarrow_i\right)={P G_0} E^\uparrow_i,\label{eq10b}
\end{eqnarray}
where $G^{\uparrow(\downarrow)}$ is the longitudinal electrical conductivity of majority (minority) electrons, $G_0=G^\uparrow +G^\downarrow $ is the total charge conductivity, and $P=(G^{\uparrow}-G^{\downarrow})/G_0$ is the spin polarization. Notice that the spin polarization $P$ appears in the charge current instead of the spin current. This can be explained by the fact that electrons with opposiste spins (parallel or anti-parallel to $\mathbf m$) are driven by opposite electric fields [see Eq.\eqref{eq5}], thus the spin current is fully polarized regardless the value of $P$. Explicitly, the spin currents driven by the spin motive force components given in Eqs.\eqref{eq8a}-\eqref{eq8c} are 
\begin{eqnarray}
J^{m}_{s,i}=\frac{g\hbar{\mu }_B G_0}{4e^2}\left[\left({\partial }_t{\mathbf m}\times {\partial }_i{\mathbf m}\right)\cdot{\mathbf m}\right],\label{eq10a}\\
J^{R,0}_{s,i}=\frac{g{\mu }_B G_0 m_e}{2e^2\hbar }\alpha_R{\left({\partial }_t{\mathbf m}\times\hat{z}\right)}_i,\label{eq10b}\\
J^{R,1}_{s,i}=\frac{g{\mu }_B G_0 m_e}{2e^2\hbar }{{{\partial }_t\alpha }_R\left({\mathbf m}\times\hat{z}\right)}_i.\label{eq10c}
\end{eqnarray}
The corresponding charge currents can be obtained by using the relation $J_e= P \frac{2e}{g\mu_B} J_s$.

In the absence of spin relaxation, the spin current $J_s$ induces a torque on the magnetization expressed as\cite{PhysRevLett.102.086601}
\begin{equation}\label{eq11}
{\mathbf T}({\mathbf J}_s)=\frac{g\mu_B}{2M_s}\partial_t{\mathbf n}_s+\frac1{M_s}\sum_i\partial_i (J_{s,i}{\mathbf m}),
\end{equation}
where ${\mathbf n}_s$ is the spin density of the conduction electrons and $M_s$ is the saturation magnetization. In the above, the total spin current can include externally supplied sources and SMF induced sources. In our case, to examine the feedback effect by SMF, we assume that there is no externally supplied spin current, {\it i.e.}, $\mathbf J_s$ is induced internally by $\mathbf m$. Recall that in the adiabatic limit the spin is alighned along the magnetization, {\it i.e.}, ${\hat{\sigma}}={\mathbf m}$, hence, the spin density ${\mathbf n}_s$ is also parallel to ${\mathbf m}$. Therefore, the torque due to the first term in Eq.~\eqref{eq11}, $\partial_t{\mathbf n}_s\propto ({\mathbf m}\times {\mathbf n}_s)$, can be neglected. The divergence of the spin current in Eq.~\eqref{eq11} can be decomposed as $\partial_i (J_{s,i}{\mathbf m})\propto(\partial_i J_{s,i}) {\mathbf m}+ J_{s,i} \partial_i{\mathbf m}$, where the first term will give rise to a parallel torque which is not interesting as it does not contribute to magnetization switching. Therefore, the torque on the magnetization in Eq.~\eqref{eq11} is simply given as
\begin{equation}\label{eq11a1}
{\mathbf T}^{ad}({\mathbf J}_s)=\frac1{M_s}({\mathbf J}_{s}\cdot\nabla){\mathbf m}.
\end{equation}

Similarly, the charge current induces a field-like spin torque in the presence of Rashba SOC, which does not involve the spin transfer mechanism and just depends on the intrinsic band structure.\cite{PhysRevB.78.212405,PhysRevB.79.094422, PhysRevB.77.214429,PhysRevB.80.094424, Tan07,Tan11} Explicitly, the field-like torque is calculated as
\begin{equation}\label{eq11a2}
{\mathbf T}^{field}=\frac{e^2}{M_s\mu_B^2m_e}\alpha_R {\mathbf m}\times({\hat z}\times {\mathbf J}_s).
\end{equation}

In the presence of the spin torque, the LLG equation is modified accordingly to become:
\begin{equation}\label{eq11b}
\frac{\partial {\mathbf m}}{\partial t}=-\gamma {\mathbf m}\times{\mathbf H}_{eff}+\gamma_G {\mathbf m}\times\frac{\partial {\mathbf m}}{\partial t}+ {\mathbf T}({\mathbf J}_s),
\end{equation}
where the second term indicates the damping torque that includes all contributions other than the SMF contribution, with  $\gamma_G$ being the intrinsic Gilbert damping constant.

\subsection{Spin torque in a static Rashba SOC}
As a pedagogical example, we first consider the spin torque due to the spin current generated by the conventional spin motive force given by Eq.~\eqref{eq10a}. Substituting the spin current expression of \eqref{eq10a} into Eqs.~\eqref{eq11} or \eqref{eq11a1}, the spin torque reads as
\begin{equation}\label{eq12b}
{\mathbf T}\left({\mathbf J}_s^m\right)=-\eta\sum_i{\left[\left({\partial }_t{\mathbf m}\times {\partial }_i{\mathbf m}\right)\cdot{\mathbf m}\rm \right]\partial_i{\mathbf m}},
\end{equation}
with $\eta=\frac{\hbar g{\mu }_B{G_0}}{4e^2M_s}$. To express the above torque in the conventional form, {\it i.e.}, ${\mathbf T}\left({{\mathbf J}}_s\right)\propto {\mathbf m}\times{\mathbf H}$, we can use the identity ${\partial }_i{\mathbf m}{\mathbf =}{\mathbf -}{\mathbf m}{\mathbf \times }\left({\mathbf m}\times {\partial }_i{\mathbf m}\right)$. With this, Eq.~\eqref{eq12b} becomes
\begin{eqnarray}
{\mathbf T}\left({{\mathbf J}}^m_s\right)={\mathbf m}\times D\partial_t{\mathbf m}\label{eq12d},
\end{eqnarray}
in which $D$ is the damping tensor given by
\begin{equation}\label{eq12e}
  D_{uv}=\eta\sum_i X_{iu}X_{iv},
\end{equation}
with $X_{iu}=\left({\mathbf m}\times {\partial }_i{\mathbf m}\right)_u$.
Eqs. \eqref{eq12d} and \eqref{eq12e} are consistent with the results of Ref. [\onlinecite{PhysRevLett.98.246601}] which considered the SMF in the absence of the spin-orbit coupling.

Similarly, in the presence of the Rashba SOC, the STT is modified to be\cite{Kim12}
\begin{eqnarray}
{\mathbf T}\left({{\mathbf J}}^m_s+{{\mathbf J}}^{R,0}_s\right)={\mathbf m}\times {\tilde D}\partial_t{\mathbf m}\label{eq13a},
\end{eqnarray}
\begin{equation}\label{eq13}
  {\tilde D}_{uv}=\eta\sum_i (X_{iu}-A_R\epsilon_{3iu})(X_{iv}-A_R\epsilon_{3iv}),
\end{equation}
where $A_R=2\alpha_Rm_e/\hbar^2$.

\subsection{Spin torque in a time-dependent Rashba SOC}
As the Rashba coupling becomes time-dependent, there is an additional SMF-induced spin current given in Eq.~\eqref{eq10c}. By substituting Eq.~\eqref{eq10c} into Eq.~\eqref{eq11a1} and Eq.~\eqref{eq11a2}, the total STT due to the dynamics of Rashba coupling is directly calculated as
\begin{equation}\label{eq14}
{{\mathbf T}}\left({{\mathbf J}}^{R,1}_s\right)=-{\mathbf m}\times \eta\partial_t A_R\sum_{iv}\left (X_{iu}-A_R \epsilon_{3iu}\right)\epsilon_{3iv}m_v.
\end{equation}

\begin{figure}
\centering
 \includegraphics[width=0.45\textwidth]{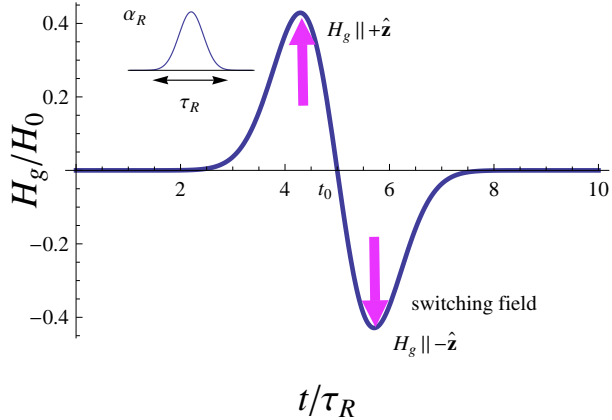}
\caption{Effective magnetic field induced by a Gaussian gate voltage with the pulse width $\tau_R$. At time $t<t_0$, the field is parallel to the magnetization, yielding no effect. At time $t>t_0$, the field is anti-parallel to the magnetization, which may result in a switching.}
\label{FigHg}
\end{figure}

To examine the nature of the torque in the presence of time-dependent Rashba SOC, we consider a uniform and static magnetization pattern. In this case, the torque on the magnetization only comes from the spin current ${\mathbf J}_s^{R1}$. Thus, the total torque is ${\mathbf{T}}=-\eta A_R\partial_t A_R ({\mathbf{m}}\times\hat z)m_z$. If we define a gate-controlled effective magnetic field via ${\mathbf T}=-\gamma\mu_0 ({\mathbf m}\times {\mathbf H}_g)$, we have
\begin{equation}\label{eq15}
{\mathbf H}_g=\frac{\eta}{\gamma\mu_0} A_R\partial_t A_R m_z {\hat z}.
\end{equation}
This field is directed along the $z-$ direction and it is generally dependent on time due to the varying $\alpha_R$. To see the effect of this field on the magnetization, we assume that initially the magnetization is in $+z$-direction, {\it i.e.}, $m_z=+1$. If at time $t=t_0$, the Rashba SOC is modulated by a Gaussian pulse such that $\alpha_R=\alpha_0 {\exp{(-\frac{(t-t_0)^2}{2\tau_R^2})}}$, where $\tau_R$ is the pulse width, the field in Eq.~\eqref{eq15} becomes  $H_g=-H_0 e^{-(t - t_0)^2/\tau_R^2} (t - t_0)/\tau_R$, where $H_0=\frac{4\eta  m_e^2\alpha_0^2}{\hbar^4\gamma\mu_0\tau_R}$ . In Fig. \ref{FigHg}, the magnetic field is illustrated during the action of the pulse. At $t<t_0$, the magnetic field is parallel to the magnetization ($+\hat z$), thus yielding no effect. However, as $t>t_0$, the field reverses to the anti-parallel direction ($-\hat z$), and it can switch the magnetization (note that in practice, there is a slight misalignment of $\mathbf m$ to the $z$-direction, which would provide the initial torque for switching). The anti-parallel field reaches its maximum value ${\mathrm {max}}|H_g|=H_0/\sqrt{2e}$ at $\Delta t=t-t_0=\tau_R/\sqrt{2}$. As an example, we consider a system with $\eta=0.2 ~{\mathrm {nm^2}}$, $\alpha_0=10^{-10}~ {\mathrm {eV.m}}$, $\tau_R=0.1~ {\mathrm {ns}}$, the switching field is estimated to be $H_g=5.6\times 10^4 {\mathrm {A/m}}$. For comparison, the switching fields in Fe, Ni, and Co are $4.5\times 10^4~ {\mathrm {A/m}}$, $1.85\times 10^4 ~{\mathrm {A/m}}$, and $59.1\times 10^4~ {\mathrm {A/m}}$, respectively.\cite{Tan11} Therefore, the modulation of Rashba SOC by gate voltage pulses is an all-electrical method for the magnetization switching, which has recently attracted many research works. \cite{PhysRevB.78.180401,Vol.pulse11, Vol.pulse14}

\section{Summary}
In summary, we proposed the enhancement and rectification of the spin-motive force in magnetization systems with Rashba SOC by application of an AC gate voltage to modulate the Rashba coupling strength. The amplitude of the SMF increases as the frequency of the sinusoidal gate voltage increases, and would exceed the conventional (static) RSOC-induced SMF if the modulation frequency is tuned to be larger than the precession frequency of the magnetization. Moreover, the AC spin-motive force can be rectified by applying a square-wave gate voltage at the resonant frequency. We also calculated the spin current induced by the SMF and the associated spin torque. We showed that the modulation of Rashba coupling by gate voltages can generate a spin-torque on a uniform and static magnetization, which can be utilized as an all-electrical method for magnetization switching.
\begin{acknowledgments}
We thank the National Research Foundation of Singapore under the Competitive Research Program ``Non-Volatile Magnetic Logic And Memory Integrated Circuit Devices" NRF-CRP9-2011-01 and ``Next Generation Spin Torque Memories" NRF-CRP12-2013-01 for financial support.
\end{acknowledgments}

%

\end{document}